\documentclass[12pt,epsfig,citesort,svgnames,dvipsnames]{article}
\usepackage{multicol}
\usepackage[all]{xy}
\usepackage{graphicx}
\usepackage{todonotes}
\usepackage{amsmath}
\usepackage{amssymb}
\usepackage{amsfonts}
\usepackage{amsthm}

\theoremstyle{definition}

\theoremstyle{remark}

\usepackage{amscd}
\usepackage{latexsym}
\usepackage[official]{eurosym}
\usepackage[english]{babel}
\usepackage[T1]{fontenc}
\usepackage[utf8]{inputenc}
\usepackage{authblk}
\usepackage{natbib}
\usepackage{xcolor}
\usepackage{hyperref}
\usepackage[nottoc]{tocbibind}
\hypersetup{
colorlinks=true,
citecolor=WildStrawberry,
anchorcolor=Black,
filecolor=cyan,
menucolor=magenta,
runcolor=cyan,
linkcolor=RoyalBlue,
linktoc=page, 
urlcolor=TealBlue}

\title{\bf {Pure shape dynamics, self-subsisting structures, and the nature of time}}
\author[a]{\normalsize Antonio Vassallo
\thanks{\href{mailto:antonio.vassallo1977@gmail.com}{antonio.vassallo1977@gmail.com}}}
\author[b]{\normalsize Pedro Naranjo\thanks{\href{mailto:pnpfisica@gmail.com}{pnpfisica@gmail.com}}}
\affil[a]{\normalsize \emph{Warsaw University of Technology, Faculty of Administration and Social Sciences, Plac Politechniki 1, 00-661 Warsaw, Poland}}
\affil[b]{\normalsize \emph{Plaza Mayor, 4/1B, 09003 Burgos, Spain}}

\date{}

\begin{document}
\maketitle

\begin{center}
Forthcoming in the special issue ``On time in the foundations of physics'' of the \emph{Journal for General Philosophy of Science}.
\end{center}

\begin{abstract}
The paper discusses the possible implications of the relational framework of Pure Shape Dynamics for the metaphysics of time. The starting point of the analysis is an interpretation of shapes in ontic structural realist terms, which gives rise to the notion of \emph{self-subsisting structure}. The relational version of a Newtonian-particle toy model is introduced and discussed as a concrete example. \\
    \\
\textbf{Keywords}: Self-subsisting structure; pure shape dynamics; materialism; reduction; time; ephemeris equation; bridge law.
\end{abstract}

\tableofcontents

\section{Introduction: A Leibnizian Argument against Leibnizian Time}\label{sec:int}

There is an obvious yet intriguing similarity between the philosophy of time and the philosophy of mind: A particular camp can be individuated in both fields, which claims that nothing fundamentally exists over and above matter. What we call ``passing of time'' or ``mental activity'' are just fancy ways to refer to purely material happenings, nothing more. What these two camps jointly uphold is not just physicalism---nobody would dispute that time, even if not reducible to matter, is part of the physical domain---but a stronger thesis, namely, \emph{reductive materialism}. In the case of mind, this stance roughly amounts to claiming that mental states are either identical to or functionally reducible to brain states. In the case of time, the general thesis is that time is nothing but an useful abstraction over material changes. This negative sentiment against time as a \emph{sui generis} entity is strong with relationalism, especially in the version that dates back to Leibniz. Indeed, the modern Leibnizian/Machian take on relational physics, originally championed by Julian Barbour and Bruno Bertotti \citep{83}, distinguishes two relational theses. A \emph{spatial} one according to which ``The only physically objective spatial information of a physical system is encoded in its shape, intended as its dimensionless and scale-invariant relational configuration,'' and a \emph{temporal} one, which maintains that ``temporal structures, such as chronological ordering, duration, and temporal flow, must be defined only in terms of changes in the relational configurations of physical systems'' \citep[][p.~110]{vasnarkos1}.

The temporal relationalism thesis provides a better insight into what the target of temporal reduction is. If the parts making up a physical system change their mutual arrangement, this means that they have ``moved'' in a relational sense (i.e., not with respect to an external embedding space). Hence, temporal relationalism suggests that time can be reduced to material motions. It is at this point that the similarity with the philosophy of mind debate generates a remarkable conceptual short circuit. To see this, it is sufficient to recall that one of the most influential arguments against reductive materialism in the philosophy of mind is due to Leibniz himself:

\begin{quote}
[P]erception and that which depends on it \emph{cannot be explained mechanically}, that is, by means of shapes and motions. And if we suppose that there were a machine whose structure makes it think, feel, and have perception, we could imagine it increased in size while keeping the same proportions, so that one could enter it as one does with a mill. If we were then to go around inside it, we would see only parts pushing one another,
and never anything which would explain a perception. This must therefore
be sought in the simple substance, and not in the compound or machine. [...]\\
(\citealp{leib}, p.~17, emphasis in the original)
\end{quote}

The argument, known as ``Leibniz' Mill,'' is best understood as an inexplicability thesis: There is nothing in the internal mechanical workings of the thinking machine that can readily explain its feelings or perceptions. All that is empirically observable is just the machine's material parts moving around, all of which does not seem sufficient to ground the existence of said feelings and perceptions. Leibniz' conclusion is that mechanical states are not sufficient for mental states. Thus, mind should be a \emph{sui generis} substance that acts as the source of feelings and perceptions. 

It does not take much to realize that a very similar inexplicability argument can be mounted against Leibnizian time itself. The thesis would sound like: It is mysterious how some sort of timeless relational change can ground highly structured temporal notions, such as that of the directed passage of time. It seems that a complex notion like that of time's arrow cannot be reconstructed by just looking at a bunch of snapshots representing different mutual arrangements of the material parts of a physical system. Even if we grant that time's arrow is a feature of the universe as a whole and, thus, we should seek it in a series of snapshots that encompass \emph{all} changes in the universe at once, still it is difficult to explain how this universal time is ``in step'' with all the possible local motions performed by physical subsystems that can be used as clocks. The conclusion seems to be that material states are not sufficient to ground structured temporal notions and, thus, time should be a \emph{sui generis} substance that acts as the source of a directed flow for the unfolding of the universal dynamics.

It is clear, then, that the ball is on the relationalists' court to show that a robust reconstruction of temporal concepts can in fact be carried out from a starting fundamental metaphysics that just admits spatial relations among material bodies and changes thereof. This is the aim of the present paper. Such a reconstruction will exploit the recent theoretical framework of Pure Shape Dynamics (PSD; cf. \citealp{726}), which delivers a totally intrinsic description of closed physical systems. Section \ref{sec:psd} will briefly set up the general framework and then present a very simple model based on classical gravity that exhibits the ``emergence'' of an arrow of time. Section \ref{sec:sss} will propose a materialist metaphysics that suits this technical framework and, thus, defuses the skeptic argument provided above. Some considerations on how to extend the discussion beyond classical gravity will be provided in section \ref{sec:disc}.

\section{Pure Shape Dynamics}\label{sec:psd}

This section will provide a brief account of the general framework underlying PSD's relational tenets (subsection \ref{subsec:ideas}), which shall be illustrated by the simplest relational model, i.e., the $3$-body system (subsection \ref{subsec:$3$-body}).

\subsection{Main Ideas}\label{subsec:ideas}

PSD was formulated to address a conceptual caveat existing in standard Shape Dynamics (SD), namely the need in this latter theory of a non-shape degree of freedom to generate the dynamical curve in shape space---something which is clearly at variance with a fully relational account of dynamics\footnote{In the $N$-body system, \citet{419} shows that best-matching w.r.t dilatations implies vanishing of so-called \emph{dilatational momentum}, which makes the theory consistent iff the potential has homogeneity degree of -2, rendering the model physically untenable. As a result, SD cannot be described as a geodesic theory with a best-matched metric on shape space when the full similarity group is taken into account, which brings in the non-shape parameter mentioned in the main text. In the case of dynamical geometry, the same applies, \emph{mutatis mutandis}, with the need of the group of \emph{volume-preserving} conformal transformations (\citealp{729}; see \citealp{514} for a comprehensive account and \citealp[][section 2]{vasnarkos1} for a concise conceptual description).}. PSD dispenses with this non-shape degree of freedom altogether by focusing on the intrinsic geometric properties of the curve in shape space, rendering dynamics explicitly unparametrized. The key to achieving this is the departure from the geodesic principles of standard SD: Besides a point and a direction, a further degree of freedom, $\kappa$, is introduced to measure the deviation from geodesic dynamics. Crucially, unlike the non-shape parameter in standard SD, $\kappa$ \emph{does} refer to an intrinsic property of the curve in shape space.

At the centre of PSD's account of dynamics is the equation of state of the unparametrized curve $\gamma_0$ in shape space $\mathcal{S}$, which reads: 

\begin{equation}
\begin{array}{rcl}
   dq^a&=&u^a(q^a,\alpha _I^a)\,, \\
   d\alpha _I^a &=&A _I^a(q^a,\alpha_I^a)\,.
   \end{array}
   \label{curve0}
\end{equation}

The right-hand side of \eqref{curve0} is described in terms of dimensionless and scale-invariant quantities. $q^a$ are points in shape space, that is, they represent the universal configurations of the system, $u^a$ is the unit tangent vector defined by the shape momenta $p_a$: 
\begin{equation}
    u^a\equiv g^{ab}(q)\frac{p_b}{\sqrt{g^{cd}p_cp_d}}\,,
    \label{unittangent}
\end{equation}
which allows us to define the direction $\phi ^A$ at $q^a$. In \eqref{unittangent}, $g^{ab}$ stands for the natural metric induced on shape space, which, crucially to our purposes in the philosophical analysis, measures the degree of \emph{dissimilarity} between shapes, not of distances thereof. Finally, $\alpha _I^a$ is the set of any further degrees of freedom needed to fully describe the system. Among these, one parameter definitely stands out: As already anticipated, $\kappa$ serves as a measure of the deviation of the curve from geodesic dynamics, enabling structure formation. We should like to elaborate on this. The physical interpretation of geodesics on shape space in PSD is simply inertial dynamics, which can be readily seen as follows: if \eqref{curve0} consists of just shapes $q^a$ and directions $\alpha _I^a\equiv\phi ^A$, the dynamical system closes at this stage, meaning no further degree of freedom is required to fully generate the curve in shape space. The relevance of this is made manifest by Poincaré's recurrence theorem, which states that geodesic dynamics on a compact space implies recurrent solution curves, thereby being unable to accommodate structure formation. This certainly applies to the model analysed in this paper, the $N$-body system, whose solution curves are geodesic loops within the associated shape space\footnote{The case of more general models, like dynamical geometry or the quantum, is pretty subtle in this regard: suffice to say that the relevant $\kappa$ keeps playing a key role in accounting for structure formation, regardless of the applicability of Poincaré's recurrence theorem, given the contentious status of the compactness of the relevant shape spaces in these cases.}. This readily highlights the essential role of $\kappa$: By measuring the deviation of the curve in shape space from geodesic dynamics, it allows us to sidestep Poincaré's theorem and, hence, bring forth variety and structure formation. 
For consistency, the elements in $\alpha _I^a$ must exhaust the set of all possible dimensionless and scale-invariant quantities that can be formed out of the different parameters entering a given theory.

Having laid down the essential ingredients of PSD, the next section will consider the explicit case of the simplest relational model, that of the $3$-body system, highlighting the ``emergence'' of the arrow of time.\footnote{\label{fn:1}The generalization of PSD to more realistic models, in particular dynamical geometry and the quantum realm, is an ongoing research program. In the case of quantum mechanics, \cite{747} is a promising first step.}


\subsection{$3$-body System}\label{subsec:$3$-body}

The first task to deal with is the construction of the relevant shape space of the $3$-body system. This system is particularly pedagogical, for the reduction of standard configuration space to shape space can be explicitly carried out. This is not possible for $N > 3$, because quotienting by rotations, unlike translations and scaling, cannot be performed explicitly. 

Given our insistence on implementing relational ideas \emph{\`a la} Leibniz and Mach, we shall consider the case of zero total energy $E$, total linear momentum $\mathbf{P}$ and total angular momentum $\mathbf{J}$, constraints that naturally follow from SD principles. The case of zero $\mathbf{J}$ has an immediate consequence: The $3$-body system is planar. It is a well-known result that the configuration space of the $3$-body system in this case has the topology of a sphere \citep{mont} and, thus, its associated shape space is referred to as the \emph{shape sphere} (\citealp[see][section 2.6, for a detailed account of this space]{712}; \citealp[see also][section 3]{merei}). Accordingly, a particularly suitable set of coordinates is given by the azimuthal, $\varphi$, and the polar, $\theta$, angles on the shape sphere, $q^a=\{\varphi\,,\theta\}$, such that the equator lies on the $\theta =\pi/2$ contour. Each point $q^a$ on the sphere stands for (the shape of) a triangle defined by the 3 particles.  

Once we have conveniently described the relevant shape space, we can proceed to describe the dynamics of the system. Since the system is self-gravitating, its dynamics is governed by the suitable projection of the Newtonian potential $V_N$ 

\begin{equation}
    V_N =-\frac{1}{M_T^2}\sum _{a<b}\frac{m_a\,m_b}{r_{ab}}\,,\,\,\, M_T=\sum _a m_a\,,\,\,\, r_{ab}\equiv ||\mathbf{r}_b-\mathbf{r}_a||\,,
\end{equation}

to shape space, accordingly referred to as the \emph{shape potential} 

\begin{equation}
 C_s =\ell_{\mathrm{rms}}V_N\,,    
\end{equation}

where 

\begin{equation}
  \ell_{\mathrm{rms}}=\frac{1}{M_T}\sqrt{\sum _{a<b}m_am_b r_{ab}^2}  
\end{equation}

is the \emph{root-mean-square length} and cancels any size inherent in $V_N$, thereby making $C_s$ scale-invariant, as it must be.

Far more interesting for our purposes is the negative of the shape potential, so-called \emph{complexity}, firstly introduced in \citet{706}, and which will play a pivotal role in this paper. The reason is the following: Given we shall be concerned with the incremental nature of the arrow of time, we need a function which increases overall. Remarkably enough, as will be shown below, this desired secular growth is a characteristic feature of complexity, whose direction of increase may thus be taken to define the arrow of time. 

In order to illustrate the meaning of complexity, it is convenient to introduce a second characteristic length of the $N$-body system, known as the \emph{mean-harmonic length} and defined as

\begin{equation}
\ell_{\mathrm{mhl}}^{-1}=\frac{1}{M_T^2}\sum _{a<b}\frac{m_a\,m_b}{r_{ab}}\,. 
    \end{equation} 

This lets us define a natural measure of complexity for the $N$-body system as 

\begin{equation}
    \mathsf{Com\,(q)}=\frac{\ell_{\mathrm{rms}}}{\ell_{\mathrm{mhl}}}\,.
    \label{complexity}
\end{equation}

A word is in order. The crucial property underlying our claim, to be given shortly, for the emergence of the arrow of time in the typical $N$-body system is the formation of structures, whereby asymptotically \emph{local} and \emph{stable} subsystems develop ever more constant symmetries. Among these subsystems, so-called \emph{Kepler} pairs provide reliable physical rods and clocks. This structure formation, which is associated with divergent complexity (see below), is clearly a \emph{dynamical} property of the model. Now, the definition of complexity itself, \eqref{complexity}, as a function on shape space, is generally a \emph{kinematical} structure that, in specific scenarios, may fail to capture the degree of relevant structure formation within a given shape. It is the attractor-driven behaviour of complexity, analysed below, within \emph{typical} dynamical trajectories on shape space that faithfully measures the formation of local and stable subsystems.

With this important qualification, the physically meaningful interpretation, at least so far as the emergence of the arrow of time is concerned, of the ratio \eqref{complexity} is as a measure of ``structure formation'': It is the ratio of the mean large-scale separations over the short-scale ones and, thus, becomes ever larger as the structure becomes ever more clumpy and clustered. 

So far, we have described the general $N$-body system for arbitrary $N$. Next, let us turn our attention to the $3$-body system. Given the representation of the shape sphere above, the explicit expression of the complexity of the $3$-body system reads \citep{merei}:

\begin{equation}
\label{complexity3}
  \mathsf{Com\,(\theta, \varphi)}=\frac{1}{M^{5/2}_T}
\sum _{a<b}\frac{(m_am_b)^{3/2}(m_a+m_b)^{-1/2}}{\sqrt{1-\sin\theta\cos(\varphi -\varphi _{ab})}}\,,  
\end{equation}

where the azimuthal angles $\varphi _{ab}$ stand for the longitudes on the equator of the shape sphere of the two-body collisions between particles $a$ and $b$.

Having suitably characterized the potential of the system, let us next consider the full Hamiltonian governing the dynamics. As shown in \citet{726}, it reads:

\begin{equation}
    H=\frac{1}{2R^2}(D^2+g^{ab}p_ap_b)+\beta\frac{C_s}{R}\equiv 0\,,
    \label{Hamiltonian}
\end{equation}

which identically vanishes because of the Machian constraint of zero total energy. In \eqref{Hamiltonian}, $D\equiv q^a\cdot p_a$ is the dilatational momentum, $p_a$ stands for the shape momenta, $g^{ab}$ refers to the metric on shape space and $R$ is a scale factor. Finally, being a homogeneous function of degree $-1$ in $R$, the Newtonian potential can be factored as $V_N(R;\theta,\varphi) =\beta R^{-1}C_s (\theta,\varphi)$, with $\beta$ a coupling constant which will be absorbed into $R$ from now on. It is illuminating to rewrite \eqref{Hamiltonian} as follows:

\begin{equation}
\label{energyconstraint}
    \varepsilon ^2 +1 = 2\,\frac{\mathsf{Com\,(\theta,\varphi)}}{\kappa}\,, 
\end{equation}

where 

\begin{equation}
\label{kappa}
\kappa\equiv\frac{g^{ab}p_ap_b}{R}\,,
\end{equation}
with $g^{ab}p_ap_b\equiv p^2$, and 

\begin{equation}
\label{varepsilon}
    \varepsilon\equiv\frac{D}{p}\,.
\end{equation}
This way the Hamiltonian constraint is expressed in terms of degrees of freedom related to the curve in shape space. To see this, let us recall that $\kappa$ satisfies the following relation \citep{726}:

\begin{equation}
\label{curvature}
    K=\frac{\sqrt{g^{ab}C_{s,a}C_{s,b}}}{\kappa}\equiv
    \frac{|\nabla C_s|}{\kappa} \,,
\end{equation}
where $K$ stands for the curvature of the curve $\gamma _0$ in shape space $\mathcal{S}$. Thus, the degree of freedom $\kappa$, which is the first element of the set $\alpha _I^a$, acquires a clear geometrical interpretation: It is a measure of the deviation of the curve in shape space from geodesic dynamics, anticipated in subsection \ref{subsec:ideas}. Finally, $\varepsilon$ is straightforwardly solved for given the energy constraint \eqref{energyconstraint}, yielding

\begin{equation}
    \varepsilon = \pm \sqrt{2\,\frac{\mathsf{Com\,(\theta,\varphi)}}{\kappa}-1}\,.
\end{equation}

Notice there is no sign ambiguity in $\varepsilon$: The growth of complexity is certainly in the direction of increasing $R$ with $D>0$ for a typical initial condition. This imposes the corresponding sign of $\varepsilon$.  

We have by now gathered all ingredients to work out the explicit equation of state of the $3$-body system, which is simply the special case of \eqref{curve0} for the following degrees of freedom: The points $q^a$ on the shape sphere are represented by the angles $\{\theta ,\varphi\}$, the tangential direction $\phi ^A$ corresponds to the two unit tangent vectors 

\begin{equation*}
    u_{\theta}=\frac{p_{\theta}}{\sqrt{2(p_{\theta}^2+\sin(\theta)^{-2}p_{\varphi}^2)}}\,,\quad\quad\quad u_{\varphi}=\frac{p_{\varphi}}{\sqrt{2(p_{\theta}^2+\sin(\theta)^{-2}p_{\varphi}^2)}}\,,
\end{equation*}
where use has been made of the explicit expression of the metric in the shape sphere, given by \citet{merei}:

\begin{equation*}
    g_{ab}=\begin{pmatrix}
\frac{1}{2}\sin ^2\theta & 0 \\
0 & \frac{1}{2}
    \end{pmatrix} ,
\end{equation*}

and finally $\kappa$ is given by \eqref{kappa}. Accordingly, the equation of state becomes:

\begin{equation}
\label{eqstate}
    \begin{array}{rcl}
d\theta&=&2\,u_{\theta}\,,  \\
d\varphi&=&2\,\sin(\theta)^{-2}u_{\varphi}\,, \\
du_{\theta}&=&\frac{1}{\kappa}\left(dC_s\,u_{\theta}-\partial _{\theta}C_s\right)+\frac{1}{2}\cos\theta\sin(\theta)^{-3}u_{\varphi}^2\,,\\ du_{\varphi}&=&\frac{1}{\kappa}\left(dC_s\,u_{\varphi}-\partial _{\varphi}C_s\right)\,,\\
d\kappa &=& -\kappa\varepsilon -2\,dC_s \,,
 \end{array}
\end{equation}
with 

\begin{equation}
    dC_s =d\theta\,\partial _{\theta}C_s + d\varphi\,\partial _{\varphi}C_s = 2\,u_{\theta}\,\partial _{\theta}C_s + 2\,\sin(\theta)^{-2}u_{\varphi}\,\partial _{\varphi}C_s\,. 
\end{equation}

Consequently, the explicit relation between $\kappa$ and the curvature becomes:

\begin{equation}
    K=\frac{2}{\kappa}\left((\partial _{\theta}C_s)^2+\sin(\theta)^{-2}\,(\partial _{\varphi}C_s)^2\right)\,.
\end{equation} 

Having given the explicit equation of state of the $3$-body system, the next task, which is the purpose of this paper, is to illustrate the emergence of the \emph{global} arrow of time, defined by the growth of complexity, as well as the emergence of a \emph{local, effective} physical clock defined by a particular particle motion that is formed in the asymptotic regime. By construction, the two clocks asymptotically march in step, whereby the ratio of the durations measured by the two clocks tends to a constant asymptotically. Thus, let us first analyze the attractor-driven behavior of complexity, whose singularities yield the formation of a pair plus a singleton. This latter fact is easily seen. Recall that complexity refers to structure formation. Now, if these structures are to be used as reliable physical rods and clocks, they better be \emph{stable}.\footnote{\label{footconst}It is worth stressing that these stable structures that are formed in the general $N$-body system develop ever better conserved charges, namely energy $E_i$, linear momentum $\mathbf{P}_i$ and angular momentum $\mathbf{J}_i$ for the $i$-th structure, which must be subject to the global Machian constraint: $\sum _i E_i=\sum _i\mathbf{P}_i=\sum _i\mathbf{J}_i=0$.} In \eqref{complexity}, $\ell _{mhl}$ tends to zero with clustering, hence making complexity diverge asymptotically. In the $3$-body system, the only physically meaningful configurations that satisfy this are asymptotic binary collisions (azimuthal angles $\varphi _{ab}$ in the shape sphere; cf. \eqref{complexity3} with $\theta =\pi/2$, that is, the equator).\footnote{Triple collisions in the $3$-body system are readily discarded: Not only are they unstable, but because both $\ell _{mhl}$ and $\ell _{rms}$ vanish, complexity reaches a finite value, thereby no interesting structure formation occurs in the first place.}

These asymptotic binary collisions correspond to effective formation of a pair plus a singleton. However, this does not quite imply that the so formed pair is a \emph{Kepler} pair, namely one that exhibits Keplerian motion. This will be tackled after the argument for the attractor behavior of complexity.\footnote{We are indebted to P. Farokhi for invaluable inputs on both the attractor behavior and the spiraling dynamics that show the emergence of a physical clock defined by the Kepler pair.} In the following, we shall provide a sketch of the argument, and simply quote the final results (see appendix \ref{Attractor} for the details). 

As is well-known, in any typical Newtonian solution, $R$ and $D$ go to infinity, with $D\sim R\sim t$ asymptotically, where $t$ is Newtonian time \citep{717}. In order to cast the analysis into PSD terms, we must find the asymptotic behavior of $\varepsilon$ and $\kappa$, which in turn demands working out the corresponding behavior of $p$, which turns out to be $p\rightarrow p_0\,t^{2/3}$. Now, (\ref{kappa}, \ref{varepsilon}) yield at once $\kappa\rightarrow \kappa_0\,t^{1/3}$ and $\varepsilon\rightarrow\varepsilon_0\,t^{1/3}$. Finally, \eqref{energyconstraint} thus further implies that complexity diverges asymptotically: $\mathsf {Com}\rightarrow\mathsf {Com}_0\,t$. As we are considering typical solutions, this in turn implies that the singularities of complexity are typical global attractors of dynamics in shape space. Thus, this attractor-driven behavior of complexity may be taken to \emph{define} the arrow of time, whereby dynamics always unfolds in the direction of increase of complexity.\footnote{\label{fn:2}Clearly, this analysis hinges on whether a suitable measure of complexity exhibiting attractor-driven behavior in the associated shape space can be defined and, hence, it depends on the specific PSD model considered. As mentioned in footnote \ref{fn:1}, the investigation of quantum and dynamical geometry models of PSD is still work in progress.} This argument may be thought of as the PSD version of standard SD, which relies on the fact that so-called \emph{Janus Points} are typical and repulsive according to the Lagrange-Jacobi relation, yielding secular growth of complexity \citep{711}. However, the repulsive character of Janus Points does not imply the \emph{attractor} behavior of complexity, so the PSD version sketched here is stronger. And, as already indicated above, this also guarantees the formation of one pair plus a singleton. That this pair is in fact a Kepler pair we shall turn next.  

We will provide a semi-qualitative argument here, leaving a more rigorous one to appendix \ref{Spiraling}. Our aim here is to supply, from the PSD perspective, an account of the well-known spiraling dynamics near singularities of complexity, which underlies structure formation. In standard SD this effect stems from the topography of shape space, in particular the infinitely deep wells of the shape potential \citep[cf.][figure 5]{712}. Clearly enough, the PSD description certainly shares these topographic features of shape space, because the difference lies in the description of the curve in shape space, which is the very same space in both frameworks. In our case, the key is the role of $\kappa$, which, recall, is a measure of the deviation of the curve in shape space from geodesic dynamics.  

Let us begin our argument by highlighting the essential difference between geodesic and non-geodesic dynamics. Geodesic dynamics lacks structure formation because it ``evades'' singularities of complexity thanks to its ``energy,'' so to speak, to escape them and go around the entire shape space. In a modified geodesic dynamics like the one in PSD, something fundamentally different occurs, exactly because of $\kappa$---or the curvature of the curve $K$, given the relation \eqref{curvature}.

To see this, let us first analyse the asymptotic expression of complexity near its singularities, assumed at $(\theta,\varphi)=(\pi/2,0)$, that is, the equator of the shape sphere.  We are interested in the first non-zero divergent term in the expansion of complexity around the singularity. Using \eqref{complexity3} yields at once the following asymptotic expression:

\begin{equation*}
  \mathsf {Com} \propto 1/\sqrt{(\theta-\pi/2)^2 + \varphi^2}\,.  
\end{equation*}

After somewhat cumbersome calculations, one shows that 
$|\nabla\mathsf {Com}|$ turns out to be asymptotically proportional to $\mathsf {Com}^2$. 

On the other hand, the Hamiltonian constraint \eqref{energyconstraint} gives
\begin{equation*}
  \kappa = \frac{2\,\mathsf{Com}}{1+\varepsilon ^2} < 2\,\mathsf{Com}\,,  
\end{equation*}
which implies 

\begin{equation*}
  K = |\nabla \mathsf{Com}|/\kappa \propto \mathsf{Com}^2/\kappa > \mathsf{Com}/2\,.
\end{equation*}

\begin{figure}[h!]
\begin{center}
\includegraphics[scale=0.2]{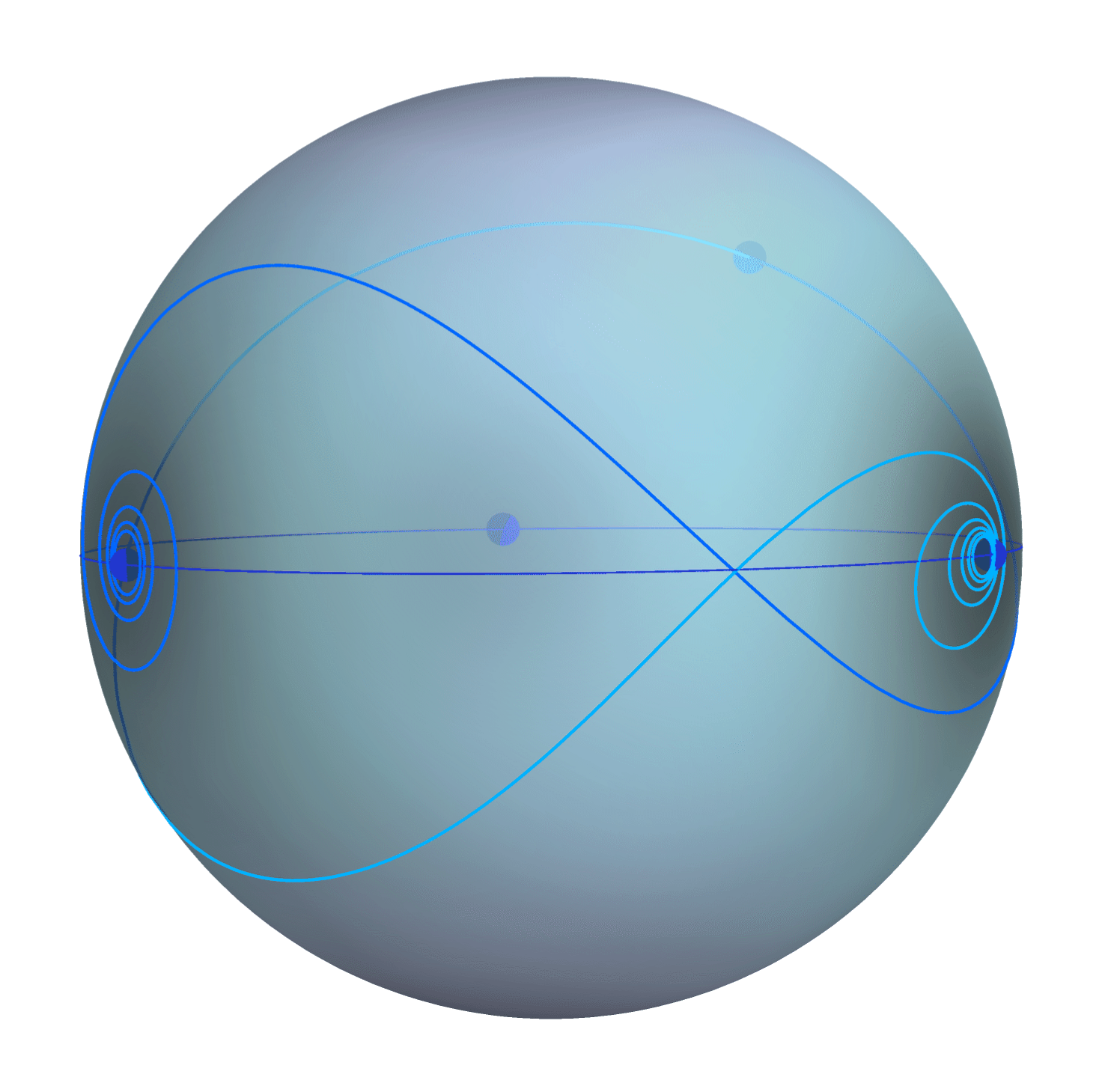}
\caption{Attractor-driven behavior of dynamics in the shape sphere. Starting at the point of absolute minimum of complexity (circle in upper half at the back) the light and dark curves show secular growth of complexity at either side, eventually exhibiting the spiraling dynamics, and ensuing Keplerian motion, near its singularities, located at the binary collisions (circles on the equator; the third binary collision of the $3$-body system is shown at the back of the shape sphere). This figure is the shape sphere counterpart of the process shown in figure \ref{fig:pair_singleton}: Crucially, whereas the Newtonian process is time-reversal symmetric, the shape sphere rendition is definitely asymmetric---the direction being given by the arrow of complexity.}
\label{fig:spiraling}
\end{center}
\end{figure}

\begin{figure}[h!]
\begin{center}
\includegraphics[scale=0.2]{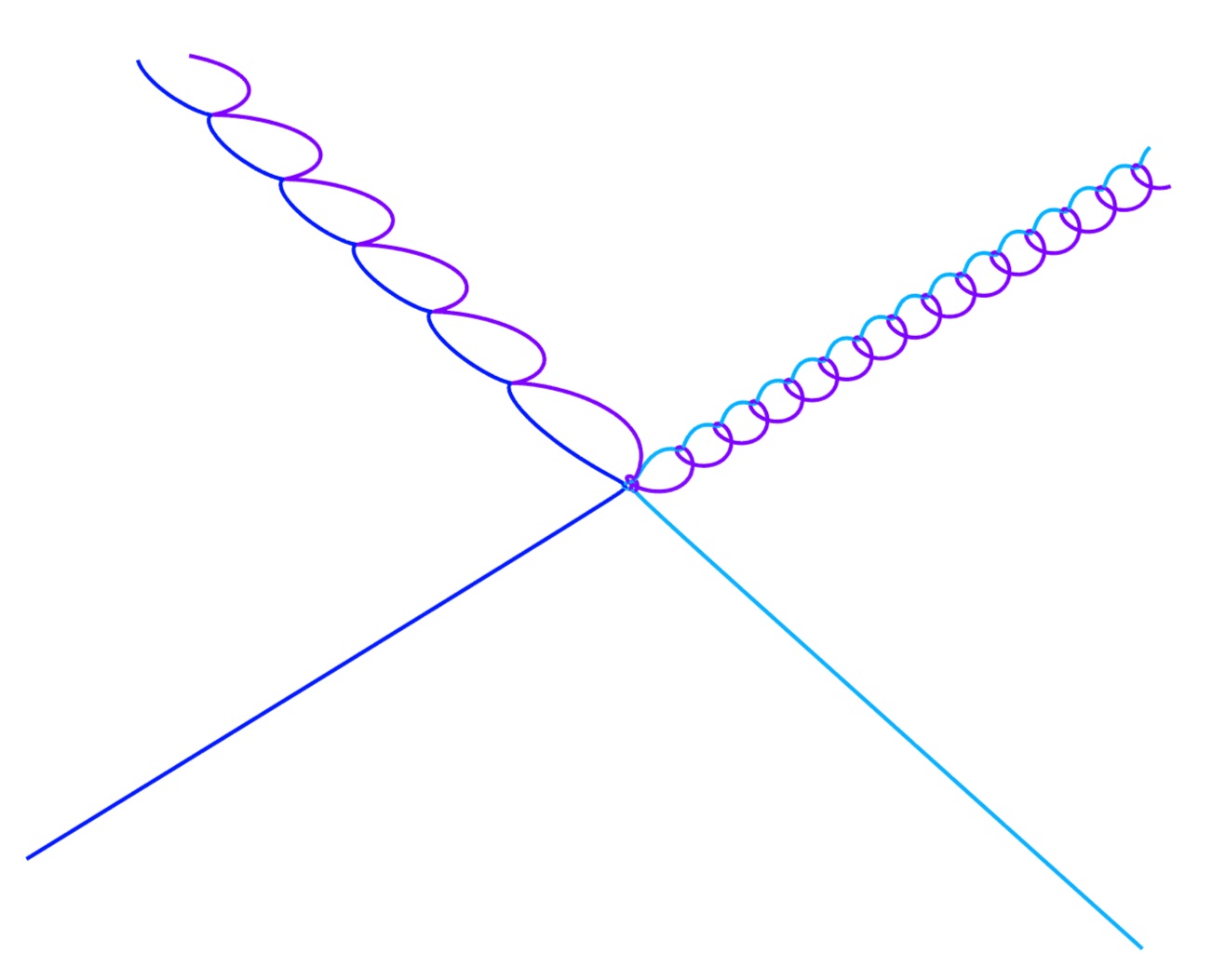}
\caption{Time-symmetric nature, stressed by lack of arrows, of Newtonian representation illustrated by the interaction between a Kepler pair and singleton. The non-trivial $3$-body interaction (centre of the diagram) develops asymptotically in both time directions, giving rise to a pair of particles that separates from the third: The singleton goes bottom left, escaping the Kepler pair, which goes upper right. By time-reversal, the previous singleton becomes part of the Kepler pair going upper left, whereas one of the two members of the previous Kepler pair now becomes a singleton going bottom right.}
\label{fig:pair_singleton}
\end{center}
\end{figure}

Given that near the singularities complexity is ever larger as the curve is closer to them, the result above shows that the curvature, too, is ever larger when approaching closer and closer the singularities, thereby causing a dramatic deviation of the curve in shape space from the geodesic one, leading it to ``get stuck'', as it were, near the attractors. This dramatic deviation underlies the spiraling behavior towards the singularities of complexity, which ultimately yields Keplerian motion: The curve in the shape sphere keeps crossing the equator, ever closer to binary collisions, \emph{periodically}, meaning the two particles are orbiting around each other (see figure \ref{fig:spiraling}; for illustrative purposes, figure \ref{fig:pair_singleton} shows the same process, but in the Newtonian representation).

The next task is to show how a physical clock can be defined by means of this Kepler pair emerging asymptotically: What this spiraling behavior implies is that complexity does not reach its singularity monotonically, but hits a \emph{local} maximum every time the curve crosses the equator (the global, absolute maximum, recall, corresponds to binary collisions). Thus, if we take the singularity of complexity as the origin, the angle subtended by the curve, which tells us how far the curve is from the equator, is a natural measure of time given by the Kepler pair. And it shares the key attribute of good old ``ephemeris time'': It is the intrinsic dynamics of the system that gets to define the time supplied by the Kepler pair. This is essentially the same idea behind the solar, or tropical, year: The time between two successive occurrences of the vernal equinox (the moment when the Sun apparently crosses the celestial equator moving north). 

It turns out there is an equation, so-called \emph{ephemeris duration equation}, which we may use to formally define a notion of Newtonian time which marches in step with the physical clock---the angle above---supplied by the Kepler pair. Given that the explicit equation for this angle is analytically unworkable, we shall take a somewhat less direct route: All we need to argue for the emergence of the arrow of time is to make the rate of change of complexity feature in the ephemeris duration equation defining the Newtonian time. As complexity is subject to attractor-driven behavior, its direction of change may be taken to define the arrow of Newtonian time. Thus, the ephemeris duration equation reads \citep{726}:

\begin{equation}
\label{ephemeris}
\frac{d}{d\,s}\log\left(\frac{ds}{dt}\right)=\frac{d}{d\,s}\log p=\frac{1}{\kappa}\left(u_{\theta}\,\partial _{\theta}\mathsf{Com}(\theta ,\varphi)+
  u_{\varphi}\,\partial _{\varphi}\mathsf{Com}(\theta ,\varphi)\right)\,,
\end{equation} 
where $s$ is the arc-length parameter and $t$ Newtonian time, and the arc-length parametrization condition, $\frac{ds}{dt}=\sqrt{g_{ab}\frac{dq^a}{dt}\frac{dq^b}{dt}}=p$, has been used. The physical interpretation of \eqref{ephemeris} is as follows: The logarithmic derivative (i) cancels the units of $p$ and (ii) yields the fraction of change of units along the curve. Given that $p$ is the ``speed'' of change of shape, the left hand side defines a duration. The right hand side follows from Hamilton's equation. The amount of Newtonian time elapsed between two configurations $q^a$ and $q^b$, call it $T$, is simply the integration of \eqref{ephemeris} between these configurations. The upshot is the following: In the asymptotic regime where the spiraling behavior unfolds and Keplerian motion develops, the local physical clock given by the angle above, which is an average measure of the rate of change of complexity, and the global Newtonian duration defined by \eqref{ephemeris}, march in step asymptotically, by construction. We speak of ``local'' physical clock because in the general $N$-body system several substructures, including Kepler pairs, get to form, each being local and effective within the whole universe. Clearly enough, in the $3$-body system, with only one Kepler pair, the physical clock defined thus happens to be the clock of the model universe.

Finally, a key point should be emphasized. Given that the Newtonian ephemeris time, \eqref{ephemeris}, contains a notion of absolute scale, it is unobservable. The true ephemeris time that local observers measure is given by the physical clock defined by the Kepler pair. The common ``passage of time'' is simply the manifestation of this Keplerian clock. 

\section{A Materialistic Metaphysics of Time}\label{sec:sss}

This section will be concerned with the reconstruction of temporal notions in the $3$-body model of PSD from a metaphysical perspective. A much general discussion of the metaphysics of PSD can be found in \citet[][section 3]{vasnarkos1}, while a detailed discussion of how a timeful description of dynamics can be achieved in the PSD framework is provided in \citet[][section 4.2]{vasnarkos2}.

\subsection{The Ontology of the $3$-Body System}

The model presented in the previous section makes it easy to see why it is reasonable to claim that PSD promotes a materialistic metaphysics (modulo the caveats expressed in footnotes \ref{fn:1} and \ref{fn:2}). The equation of state \eqref{eqstate} is cast in terms of degrees of freedom \emph{internal} to the $3$-body system, i.e., the dimensionless ratios of the inter-particle separations. Hence, at no point in the dynamical description are notions linked to an external embedding space and time drawn into the picture: It's all about the (change of) mutual arrangement of the three material particles and nothing else. Everything there is to the model is a relational configuration of bits of matter, i.e., a triangular shape. Two objections naturally arise at this point.

Firstly, it may be pointed out that, according to the model, there is not just three material particles, but also the \emph{spatial} relations holding among them. Hence, there is something else in the ontology of the model beyond pure matter. The reply to this objection is that there is nothing to reply to. Not only is it true that spatial relations prominently enter the metaphysical picture, but this \emph{reinforces} the materialistic take on the ontology. This is because the PSD framework eschews from the physical picture any ``absolute'' physical magnitude---including mass---in favor of dimensionless and scale-invariant quantities, which implies that it is not possible to physically characterize one of the particles in isolation from the other two. In other words, PSD pushes a metaphysical reading of a $3$-body shape as a structure consisting of Euclidean conformal relations holding among three \emph{relata} that, by themselves, have no intrinsic physical features. There is nothing to these \emph{relata} beyond the relations they stand in, as testified by the fact that PSD's equation of state features just shape degrees of freedom and nothing else that can be interpreted as intrinsic properties of the particles (not even their identity, if we consider that the permutation group can also be quotiented out of shape space, meaning that the particles in a configuration can be arbitrarily labelled without changing anything in the physical information carried by their shape). Hence, it is exactly by virtue of standing in these particular spatial relations that these \emph{relata} can be called ``material particles.'' This view is obviously reminiscent of the Cartesian conception of matter as \emph{res extensa}, namely, the ``stuff'' that constitutes the physical, extended, and measurable aspect of reality. In the present context, this view amounts to holding that the property of being material is just as relational as any other: This is the sense in which the spatial ideology \emph{reinforces} the materialistic metaphysics considered. Note how this whole discussion naturally leads to interpreting the relational configurations in ontic structural realist terms. The three ``particles'' making up a shape do not exist independently of each other: Each is determined and individuated by its relations to the other two. Given that shapes are intrinsically determined, without any need to place them in an external space and time, the structures they represent are called \emph{self-subsisting} \citep[cf.][section 3]{vasnarkos2}.

The second objection to this ontological reading of the model is that no physically meaningful concept of ``motion'' can be articulated in the absence of an embedding spacetime. Even granting that the equation of state describes a change in the spatial relations constituting a self-subsisting structure, still it is mysterious how to make sense of such a change without assuming that it unfolds in time---even just a weak version of it, such as that provided by a parametrization that labels the shapes ``from outside'' the system. How is it possible to say that the $3$-body system \emph{started} in a certain configuration $q_1$ and \emph{then} changed to a different one $q_2$ without assuming certain temporal stages $t_1$ and $t_2$ that act as ``slots'' that make the comparison between $q_1$ and $q_2$ possible? In other words, doesn't change presuppose some minimal temporal connotation? 

The reply to this objection is that it places too strong a constraint on the characterization of genuine change. All that is needed to have change is a pre-existing ordering, but such an ordering need not be temporal. Too see this, it is sufficient to recall that a solution of the equation of state is an unparametrized curve in shape space. Given that each point in shape space represents a possible relational configuration, we immediately see that solving the equation of state means taking the physically possible shapes according to some initial conditions and ordering them along a curve. The ``distance'' between two configurations in shape space does not measure their spatial or temporal interval, but how much they differ. Hence, the nature of the dynamical ordering underlying a solution of the equation of state is readily found: It is a topological ordering in terms of degrees of similarity. In order to make ontological sense of such an ordering, it is first of all important to notice that a curve representing a solution of the equation of state given certain contingent initial conditions contains \emph{all} the information about the dynamical development of the $3$-body system given such initial conditions. But such a system is all there is in the universe. Hence, a dynamical curve represents the entire history of relational changes in the universe given certain conditions obtain---different initial conditions will generate different histories. It is then natural to consider a dynamical curve in shape space as a physically possible world in a Lewisian sense since all there is to know about such a universe is contained in this curve. The ontological reading of a solution of the equation of state is thus the following: A dynamical curve in shape space represents a possible world according to PSD, where all the physically realized shapes are given all at once in a timeless sense, and ordered in terms of their degree of similarity. In this picture, change is just variation across a dynamical curve. Such a variation is couched in purely intrinsic terms (differences in shape as measured by the metric on shape space) and does not need any external labeling to be established.

\subsection{No Time Without Identity}

The ontology sketched above depicts a cluster of physically possible worlds---one for each particular solution of the equation of state given a physically possible set of initial conditions. Each world is a ``still life'' where all the dynamical stages of the evolution according to the equation of state (and the particular initial conditions) are actualized all at once in a timeless manner and ordered in terms of similarity (the more similar two shapes are, the ``closer''). These worlds are clearly distinct, e.g., there are shapes that exist in some worlds but not in others. From this point of view, the proposed ontology does not amount to shape space realism (e.g., like Julian Barbour's Platonia; see \citealp{10}) in that this latter stance would maintain that all conceivable shapes are actualized, irrespective of whether or not they appear in a dynamical solution. In a nutshell, shape space realism represents a pre-dynamical ontic commitment towards shapes \emph{simpliciter}, whereas the ontology presented here is dynamically-informed in that it takes different dynamical curves to represent distinct possible worlds.

There are at least two reasons to prefer the present ontology to shape space realism. The first is that shape space realism renders the equation of state rather mysterious: If all the conceivable shapes exist at once, what is the meaning of a dynamical solution ``selecting'' a small subset of them based on the initial conditions given? No mystery, of course, arises if we instead take each dynamical curve that can be traced in shape space as the entire physical information encoded in a possible world. Secondly, and relatedly, considering shape space as a whole as a real entity washes away the linear ordering inherent in dynamical curves, thus making it much more difficult to recover temporal concepts. Instead, once the notion of physically possible world according to PSD is established, time can be readily reconstructed from the configurational history making up each of such worlds. 

The starting point of such a reconstruction is, of course, the order inherent in a sequence of shapes. Such an ordering is topological in nature: Recall that a dynamical curve is unparametrized and, hence, it lacks any physically objective means to establish a ``distance'' between two points. The only available notion is that of ``nearness,'' which means that, e.g., given three shapes $q^a$, $q^b$, and $q^c$, the most that can be said is whether $q^b$ is closer to $q^a$ than $q^c$, based on their similarity, which is measured by the metric on shape space. Such a weak topological ordering is not directed and is just grounded on a primitive notion of Leibnizian plurality: There are distinct shapes and no two distinct shapes can be identical in all relational aspects, i.e., perfectly similar.


This primitive distinctness has radical ontological overtones, bolstered by the structuralist spirit of the metaphysics at hand. This means that we are not just talking about \emph{distinct} configurations of the \emph{same} three particles, because the particles (i) have no intrinsic identity and (ii) are individuated through the relations they stand in, which are of course different from shape to shape. As a result, there is no fundamental notion of identity over configurational stages: A dynamical curve does not fundamentally depict the relational evolution of three particles, but just a weakly ordered sequence of triangular shapes. This is the radically timeless spirit of the proposed ontology: It is not even possible to claim that each shape represents a snapshot of a $3$-particle system at a certain ``moment'' of the dynamical evolution. The challenge, then, is to show how time can appear out of this crystallized fundamental realm.

\begin{figure}[h]
\begin{center}
\includegraphics[scale=0.8]{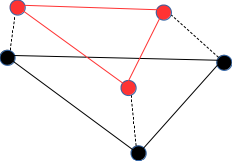}
\caption{By comparing two distinct configurations (upper and lower triangles) and identifying the quasi-overlapping particles (dotted lines) it is possible to reconstruct a dynamical picture of a $3$-particle system changing its shape.}\label{fig:juxt}
\end{center}
\end{figure}

The key to regaining a notion of identity over configurations for the particles and, with it, a temporal picture of the physical evolution resides in the notion of complexity discussed from a physical point of view in section \ref{sec:psd}. Complexity is basically a measure of clustering in the relational configuration and, as such, depends on dimensionless ratios of inter-particle separations. To each configuration in a dynamical curve a value of complexity can be assigned based on \eqref{complexity} (more precisely, \eqref{complexity3} for the $3$-body system), and this equation presupposes a labelling---and, hence, an identification---of each particle in the configuration. However, if there is no notion of identity across configurations, there is nothing that guarantees that the labeling stays consistent when moving along the curve. If that was the case, the values of the complexity function would be randomly distributed along the curve. But we already know they are not: These values tend to consistently grow secularly. This means that the behavior of complexity picks out a unique way to ``match'' the particles across configurations---i.e., give a consistent labeling along the curve. This unique way comes from the similarity ordering across a curve (encoded in the metric on shape space), which makes it easy to compare neighboring shapes by juxtaposing them and, then, identify the particles that quasi-overlap (see Figure \ref{fig:juxt}). Hence, it is straightforward to recover a notion of identity over configurations, even if this concept is not ontologically fundamental.

With the notion of identity over configurations in place, the dynamical curve can be interpreted as a sequence of dynamical stages of a unique $3$-particle system. At this point, the idea of relational motions of the particles can be easily conveyed via a ``stock film'' analogy. In the same vein as an animated picture is realized by showing a sequence of still frames ordered by their similarity, the picture of three particles swirling around is recovered by fixing a starting configuration on the curve and then traversing the curve in a continuous manner. If the complexity of the sequence of configurations increases, the traversing represents the moving forward in time: This is the sense in which complexity represents an intrinsic ``time stamp'' for each configuration. So, in the end, a dynamical curve is the ``film strip'' containing the whole dynamical evolution of a physically possible world according to equation \eqref{eqstate}, with complexity providing information about the direction in which the strip should be scanned.\footnote{Clearly, this whole analysis can be repeated for whatever model of PSD that (a) is represented by a well-behaved enough dynamical curve (e.g., a continuous one) and (b) makes it possible to construct a viable notion of complexity exhibiting attractor-driven behavior in the associated shape space. Here it is worth stressing again that, while conditions (a) and (b) are certainly fulfilled for typical $N$-body PSD models, whether the same holds for more general models is still object of ongoing investigation (see the remarks made in the conclusive section).}

It is interesting to note that, while a dynamical curve can be indefinitely traversed in the direction of increasing complexity, this does not happen in reverse, because of the spiraling behavior described in subsection \ref{subsec:$3$-body}: Whatever the traversing direction chosen, the value of complexity will grow secularly, and eventually reach the attractor behavior. In other words, the dynamical laws individuate a privileged direction for the unfolding of dynamics. Metaphysically speaking, this means that the arrow of time is ``crafted'' in the fundamental material states of affairs making up a world. There is no need for invoking fundamental temporal facts on top of material ones: Material facts are enough to do the job of grounding the dynamical description of a $3$-particle system. Speaking about time is just a concise way to speak about material happenings. This is, of course, the fulfillment of the reductive materialists' dream. 


\subsection{What Kind of Reduction?}

The question, at this point, is whether the reductionist thesis underlying the reconstruction of time sketched above is able to overcome the Leibnizian objection considered in section \ref{sec:int}. To assess this, a more detailed discussion of such a thesis is in order.

The previous sections insisted on a specific metaphysical picture: Everything there is to a physically possible world according to the equation of state is a series of \emph{material} facts encoded in the conformal structure making up shapes. This comes from the Cartesian-like definition of matter brought forward in this context, according to which ``being material'' just means being a place in such a conformal structure---i.e., being a \emph{relatum} in this web of spatial relations. Moreover, following the Leibnizian idea that no two distinct entities can share all discernible features (usually referred to as the Principle of Identity of Indiscernibles), each shape instantiates ``its own'' set of material facts that, consequently, fully define and individuate it: This is why each shape is a self-subsisting structure. The bottom line of this story is that there is nothing inherently temporal in this metaphysical picture, but this fundamental furniture is robust enough to build temporal notions on top of that. What does such a building metaphor really amount to? Loosely speaking, temporal facts are determined by material facts in such a way that the former can always be analyzed and explained in terms of the latter, but not viceversa (e.g., time moves forward because complexity grows, but it is not the case that complexity grows because time moves forward). This whole discussion backs up the idea that the dependence relation between time and matter represents a case for reduction. To get a clearer idea of how reduction works in the present case, let's start from this intuitive characterization of reduction provided by \citet{redux}:

\begin{quote}
The English verb `reduce', derives from the Latin `\emph{reducere}', whose literal meaning [is] `to bring back' [...] If one asserts that the mental reduces to the physical, that heat reduces to kinetic molecular energy, or that one theory reduces to another theory, one implies that in some relevant sense the reduced theory can be \emph{brought back} to the reducing theory, the mental can be \emph{brought back} to the physical, or heat can be \emph{brought back} to molecular kinetic energy.\\
(\emph{ibid.}, emphases in the original)
\end{quote}

It is pretty straightforward to find the above-mentioned ``relevant sense'' in the case of Newtonian time in PSD. Indeed, the ephemeris duration equation \eqref{ephemeris} is nothing but a way of ``switching'' back and forth between a temporal description of the $3$-body system's dynamics proper of Newtonian mechanics and the timeless relational one of PSD, which relies on the complexity function's rate of change. Hence, any description of three material particles moving to the tick of a universal clock can be ``brought back'' to the description of a sequence of self-subsisting structures ordered in such a way that their value of complexity grows secularly due to the attractor behavior and, in particular, the spiraling dynamics described earlier. From this point of view, equation \eqref{ephemeris} can be interpreted as a genuine Nagelian bridge law, i.e., a coordinate definition linking the reduced to the reducing notions (time and complexity, respectively; see \citealp{nag}, for the original source). 

The nature of the link between time and complexity is conceptually very strong. The ephemeris equation does not just establish how complexity explains away time; it establishes that temporal facts \emph{are} material (relational) facts. Therefore, it is natural to interpret equation \eqref{ephemeris} as establishing an analytical link---in the sense of its being conceptually \emph{a priori} rather than contingent on some empirical conditions obtaining---between time and complexity. This analytical link takes the form of an identity between the extensions of temporal and complexity predicates. In other words, what makes a predicate like ``$\Delta x$ seconds of Newtonian time have elapsed'' true is the same that makes the predicate ``The average of complexity grows by $\Delta y$'' true. 

A possible objection against this reductionist framework surfaces, at this point. Take, say, a $6$-body system solution that is basically two $3$-body systems undergoing the formation of a Kepler pair with the third particle (the singleton) receding from it with inertial motion. Each half-system would represent a good clock which, as we have seen in section \ref{sec:psd}, would march, asymptotically with ever better accuracy, in step with the other (recall, the ratio of the durations measured by each Kepler pair would tend to a constant asymptotically), thus giving the impression of measuring a universal Newtonian time $t$. Take now a predicate about $t$, call it $P$: The extension $Ext(P)$ of this predicate is identical to the extension $Ext(p_{1})$ of some predicate $p_1$ consisting of a subset of facts about one of the two Kepler pairs. This is obvious, given that this Kepler pair measures $t$. However, also the second Kepler pair measures $t$, which means that there is going to be another predicate $p_2$ about this latter pair, such that $Ext(p_{2})=Ext(P)$. But, if this is the case, things do not add up anymore given that $Ext(p_{2}) \neq Ext(p_{1})$, since the facts about the two Kepler pairs are distinct. This objection should sound very familiar to the reader versed with the philosophy of mind literature, since it is just an iteration of the multiple realizability argument, which dates back to \citet{put}. 

This multiple realizability argument was originally devised to attack the idea that mental states are identical to brain states but, given the close similarity that the present discussion has with reductionist theories of mind, it applies almost \emph{verbatim} to this context. Why ``almost''? Because, unlike the mind/brain identity theory, in this context there is a quick way out of this impasse that exploits the very nature of self-subsisting structures. In a nutshell, the facts making $p_1$ and $p_2$ true do not involve anything regarding the intrinsic identity of the particles making up the two $3$-body systems, nor any other intrinsic property thereof, which means that they are purely structural. Hence, there is no problem with claiming that these facts \emph{are the same} for both systems: What makes $p_1$ and $p_2$ true is a certain relational arrangement that both systems can instantiate at once, which means $Ext(p_{1}) = Ext(p_{2})$. Having embraced Leibniz' Principle of the Identity of Indiscernibles, this move is not only legitimate but also necessary to keep the framework consistent. Does this mean that the two systems ``collapse'' into one due to their indiscernibility? Of course not, because the facts making $p_1$ and $p_2$ true are not the only ones inherent into the two subsystems and the $6$-body structure as a whole. 

For example, there are also relational facts that ground the emergent notion of local energy of each subsystem. Hence, if the two subsystems are predicated of having different energies, the underlying material facts that ground this predication will be structurally different as well. This also applies to the other asymptotically developed charges, i.e., linear and angular momentum. Even if the two sets of local charges must satisfy the global Machian constraint of zero total energy, linear and angular momentum, they are different nonetheless: In the case of the 6-body system, the global Machian constraint amounts to the simple relation $E_1=-E_2\,,\mathbf{J}_1=-\mathbf{J}_2\,,\mathbf{P}_1=-\mathbf{P}_2$ for the respective local charges of the two Kepler pairs and, in the general $N$-body system, the differences between two Kepler pairs arbitrarily chosen are even more clear-cut (cf. footnote \ref{footconst}). Moreover, on top of all the previous mentioned facts, there will also be facts about the mutual arrangement of the two halves of the system that will make it possible to tell the two Kepler pairs apart. This structuralist reply has the added value of beautifully accounting for the overlap of the global Newtonian time with the local clocks represented by the Kepler pairs: Structurally speaking, they are one and the same thing, so there is nothing mysterious in the global and the local perspectives being in perfect accordance.

It is by now abundantly clear that the type of reduction involved in this context is essentially ontological in that it specifies the kind of ontic commitments involved when using temporal concepts in PSD. Therefore, it might seem straightforward to take the present metaphysical framework not only as reductionist, but even \emph{eliminativist} towards time. Instead, this turns out to be a contentious point. This is because saying that time is totally eschewed from the picture would mean not only that temporal concepts are just a concise way to refer to material facts, but also that these concepts are, in the end, cognitively useless and, as such, bear no explanatory value whatsoever. Think of the notion of caloric, which is an empirically inaccurate way to describe thermal phenomena and, as such, has been totally eliminated in the context of modern statistical mechanics. By the same token, we should reject explanations involving temporal concepts as misleading. Such an eliminativist reading, however, sounds too radical, since---unlike caloric-based explanations---explanations involving temporal concepts seem to work pretty well both in scientific practice and everyday life. Thankfully, the structuralist spirit of the proposed metaphysics comes to the rescue again. 

Indeed, this structuralist spirit maintains that the empirically adequate aspects of the world captured by a theory are relational and, hence, they are carried over untouched through theory change. Paraphrasing the famous example by \citet[][p.~15]{675}, what once was called ``electron's motion'' in classical electromagnetism has later become ``electron's propagator'' in quantum electrodynamics: What the first term got empirically right is the same referent of the latter. Given that this referent is not a ``thing'' but, rather, a set of relational facts, there is nothing mysterious in claiming that both terms ``point'' at the same aspects of physical reality. Therefore, it is not cognitively useless or misleading to explain certain phenomena in terms of electrons' motion (e.g., the filling of valence bands in semiconductors), even if such a jargon is not the one adopted in our current best theory of electromagnetism. By the same token, it is perfectly fine to keep temporal notions in place and use them in physical explanations as long as it is clear that such notions do not refer to the properties of a \emph{sui generis} substance, but to fundamental relational facts involving material structures. In conclusion, the proposed metaphysics surely eliminates in an ontological sense time as a substance, but, from an epistemic perspective, it just reduces temporal concepts to material ones, meaning that time remains a cognitively useful concept.

With this last remark in place, it is finally apparent that the reconstruction of time carried out in the previous paragraphs escapes the charge of inexplicability put forward in section \ref{sec:int}. Self-subsisting structures and their similarity-based dynamical ordering are sufficient to reconstruct temporal notions from a purely material ontological bedrock. The key to achieving this, as we have just seen, is to stress that what is to be reconstructed is not a ``thing'' called time, but the way temporal parlance picks up empirically adequate referents, with these referents being material facts about self-subsisting structures. All this information is encoded in the ephemeris equation \eqref{ephemeris}, which serves as a Rosetta Stone that links temporal terms to material ones.






\section{Conclusion}\label{sec:disc}

We have illustrated the role of the attractor behavior of complexity as the defining physical mechanism underlying the arrow of time in the simplest relational model, the $3$-body system. This has allowed us to provide explicit equations, but the main claim easily extends to arbitrary $N$. What about more realistic models, ones encoding the dynamical nature of geometry, as in general relativity (GR), or the quantum? Let us make some general comments on these generalisations. 

Regarding dynamical geometry, one may consider two broad categories. First, a possible extension involves a suitable measure of complexity in the standard formulation of dynamical geometry, namely finding the relevant counterpart of \eqref{complexity} when the role of dynamical geometry is taken into account. A first proposal has been put forward in \citet[section 3.6]{712} in the context of \emph{vacuum} Bianchi IX cosmological model. We have stressed ``vacuum'' because this tentative measure of complexity conceptually differs from that of the $N$-body system, which, by construction, measures the complexity of \emph{matter} degrees of freedom.\footnote{A straightforward extension to a measure of complexity including also matter degrees of freedom may be put forward, leaving technicalities aside, but this will contain dynamical geometry as well.} This leads us to the second path, which tries to remain loyal to Leibnizian relationalism: Whatever measure of complexity happens to suit the empirical content of GR-like views, it should be sought \emph{within} matter degrees of freedom. This means that, regardless of the geometrical framework, the utterly unobservable character of the purely vacuum geometry should make us focus on its observable, dynamical influence on the spatial relations between these matter degrees of freedom. This is definitely a bold proposal, both conceptually and mathematically: Not only does it undermine Einstein's great legacy, whereby space(time) is elevated to the rank of a dynamical character, standing on equal footing with the rest of physical degrees of freedom, but it is certainly unclear how a theory without dynamical geometry \emph{and} empirically equivalent to GR may be built. This is the subject of ongoing research.

As for the extension of the framework to the quantum, the fact the relational framework of SD, broadly construed, has yet to accommodate quantum phenomena clearly poses an issue. Things become somewhat simpler if one adopts a formulation of quantum theory that shares the ontological traits of the $N$-body system. This is precisely what has been proposed in \citep{747}, where the PSD version of a de Broglie-Bohm model of the $N$-body system is studied. Remarkably, the numerical analysis shows, at least for the $3$-body model, the expected classical behavior for the universe asymptotically. Although quantum dynamics does affect the evolution of the early Universe, the crucial attractor-driven behavior of complexity, and, with it, the ensuing structure formation, remain. However, it is still not completely clear under which specific physical conditions subsystems satisfying Born statistics emerge. Also this is subject of ongoing research.

From a metaphysical standpoint, an important part of the story the reductive materialist should provide about the appearance of time out of a fundamentally timeless reality concerns how living beings like us can get to perceive the flow of time. This is where the analogy between philosophy of mind and philosophy of time hinted at in the beginning becomes a straight overlap. Indeed, the first point to address when investigating the issue is how \emph{minds} can emerge from material self-subsisting structures. An educated guess in this direction is that, also in this case, the notion of complexity is going to play a major role. This is because complexity provides physical information regarding (i) the formation of subsystems in the universe---biological brains being among those---and (ii) the amount of structure inherent in these subsystems. Moreover, it is to expect that brains will be richly structured subsystems capable of interacting with the ``environment'' (i.e., other subsystems) and, as a result of this interaction, storing a huge amount of information in the form of records of the past. Clearly, this characterization is not sufficient to grant the emergence of full-fledged minds (in the sense of brain processes unfolding in time), but it is plausibly necessary for that. Hence, a first step in this research direction would be to investigate in further detail the subsystem formation mechanism underlying more physically realistic models of PSD, which leads back to the question about how to generalize the notion of complexity beyond classical mechanics. Will emergent minds in this picture (provided such a reconstruction is achievable) stand a chance against Leibniz' Mill argument? The answers to this and the other questions just considered will hopefully come, \emph{in time}. 


\appendix

\section{Attractor-driven behavior of complexity}\label{Attractor} 

We shall give the steps leading to the asymptotic behavior of shape momenta. Theorem 1 in \citet{717} and the typical behavior $R\sim t$ lead to the following asymptotic expression for Newtonian positions:

\begin{equation*}
    r_{i_{cm}}^{\alpha}\rightarrow r_{i_0}^{\alpha}\,t + f_i^{\alpha}(t)\,t^{2/3}\,,\quad \quad \forall i=1,2,\ldots ,N\,,\quad \quad \alpha=1,2,3\,,
\end{equation*} 
where $f_i^{\alpha}(t)\rightarrow$ constant as $t\rightarrow\infty$. In order to cast this in relational terms, recall that shape variables are, by definition, ratios of inter-particle separations, so there must exist some locally analytic function such that 
\begin{equation*}
 q^a:= F\left(\frac{r_{i_{cm}}^{\alpha}}{r_{j_{cm}}^{\alpha}},\cdots\right) + \cdots    
\end{equation*}
 
Thus, we have 

\begin{equation*}
    q^a\rightarrow F\left(\frac{|r_{i_0}^{\alpha}\,t + f_i^{\alpha}(t)\,t^{2/3}|}{|r_{j_0}^{\alpha}\,t + f_j^{\alpha}(t)\,t^{2/3}|},\cdots\right) =F\left(\frac{|r_{i_0}^{\alpha} + f_i^{\alpha}(t)\,t^{-1/3}|}{|r_{j_0}^{\alpha} + f_j^{\alpha}(t)\,t^{-1/3}|},\cdots\right) \rightarrow F_0 + F_{00}\,f(t)\,t^{-1/3}\,,
\end{equation*}
where the constants $F_0$ and $F_{00}$ depend on $F\left(\frac{|r_{i_0}^{\alpha}|}{|r_{j_0}^{\alpha}|},\cdots\right)$ and its first derivative, respectively. Therefore,

\begin{equation*}
    \Dot{q}^a\rightarrow -\frac{1}{3}F_{00}\,f(t)\,t^{-4/3}\,,
\end{equation*} 
as the term involving $\Dot{f}(t)$ diverges slower. Next, given \eqref{Hamiltonian}, Hamilton's equation readily yields

\begin{equation*}
    \Dot{q}^a =\left\{q^a,\frac{p^2}{2\,R^2}\right\}=\frac{p}{R^2}\,,
\end{equation*}
which finally leads to the asymptotic behavior of shape momenta: $p\rightarrow p_0\,t^{2/3}$, recalling $f(t)\rightarrow$ constant asymptotically and $R\sim t$. 

\section{Spiraling}
\label{Spiraling} 

Let us provide a more complete analysis of the spiraling behavior and show that it in fact leads to the singularities of complexity. The argument assumes the curve is already subject to the attractor behavior of complexity, which is already established in the main body (section \ref{subsec:$3$-body}).

Let $u$ be the direction of the curve near singularities, where $\kappa$ and $\mathsf{Com}$ are both close to infinity. By definition, then, $du := du/ds$ is the acceleration vector. Let $\xi := du - u$. Clearly, $\xi$ measures the deviation of the acceleration from the direction along which the curve is already going. Next, let us look at the direction of $\xi$. If it points towards the singularities, it means the curve bends around them, namely, as it is subject to the attractor behavior, it spirals around the attractors. 

Near the singularity ($\varphi = 0\,,\theta =\pi/2$), the expression of complexity reads:

\begin{equation*}
 \mathsf{Com}(\varphi,\theta) = \frac{\mathsf{Com}_0}{\sqrt{\varphi^2 + (\theta-\pi/2)^2}}\,.   
\end{equation*}

A direct calculation shows that 

\begin{equation*}
du := (du_\varphi,du_\theta) = -\frac{1}{\kappa}\frac{\mathsf{Com}^3}{\mathsf{Com}_0^2} (\varphi,\theta) - \frac{1}{\kappa}\frac{d\,\mathsf{Com}}{ds}(u_\varphi,u_\theta)\,.
\end{equation*} 

So

\begin{equation*}
 \xi = -\frac{1}{\kappa}\frac{\mathsf{Com}^3}{\mathsf{Com}_0^2} (\varphi,\theta) - \left(\frac{1}{\kappa}\frac{d\,\mathsf{Com}}{ds} + 1\right) (u_\varphi,u_\theta)\,.    
\end{equation*}

Next, let us determine the direction of $\xi$. As the last term, $-(u_\varphi,u_\theta)$, is of order 1 in magnitude, let us focus on the other two terms: $\frac{\mathsf{Com}^3}{\mathsf{Com}_0^2} (\varphi,\theta)$ and $\frac{d\,\mathsf{Com}}{ds}(u_\varphi,u_\theta)$.

We have

\begin{equation*}
 \left|\frac{d\,\mathsf{Com}}{ds}\right| = \frac{\mathsf{Com}^3}{\mathsf{Com}_0^2} \left|(\varphi,\theta)\cdot(u_\varphi,u_\theta)\right|\leq  \frac{\mathsf{Com}^3}{\mathsf{Com}_0^2} \sqrt{2} \sqrt{\varphi^2 + (\theta-\pi/2)^2} = \sqrt{2} \frac{\mathsf{Com}^2}{\mathsf{Com}_0}\,,   
\end{equation*}

where the inequality comes from the Cauchy-Schwarz inequality. And 

\begin{equation*}
 \left|\frac{\mathsf{Com}^3}{\mathsf{Com}_0^2} (\varphi,\theta)\right| = \frac{\mathsf{Com}^3}{\mathsf{Com}_0^2} \sqrt{2} \sqrt{\varphi^2 + (\theta-\pi/2)^2} = \sqrt{2} \frac{\mathsf{Com}^2}{\mathsf{Com}_0}\,.   
\end{equation*} 

Now, the equality in the Cauchy-Schwarz inequality does not hold, given we are assuming $u$ is not already in the direction of singularities, namely we are considering typical cases and, hence, excluding special solutions which terminate at the singularity without exhibiting any Keplerian motion. Thus, the first term in $\xi$ is larger than the second one, meaning that the dominant term in the direction is along $-(\varphi,\theta)$, i.e., pointing towards the singularity, implying the curve spirals around the attractors. 

\pdfbookmark[1]{Acknowledgements}{acknowledgements}
\begin{center}
\textbf{Acknowledgements}:
\end{center}
We thank the editors of this special issue for inviting us to contribute. Many thanks also to an anonymous referee for his constructive criticism. Finally, we gratefully acknowledge financial support from the Polish National Science Centre, grant No. 2019/33/B/HS1/01772.

\bibliography{biblio}

\end{document}